\documentstyle[epsf]{elsart}

\begin{document}
\begin{frontmatter}
\title{Formation and decay of super heavy systems}

\author[LNS,JAERI]{Toshiki Maruyama} \ead{maru@hadron02.tokai.jaeri.go.jp}
\author[LNS]{Aldo Bonasera} \ead{bonasera@lns.infn.it} 
\author[LNS,CT]{Massimo Papa} \ead{papa@lns.infn.it}
\author[JAERI]{Satoshi Chiba} \ead{chiba@hadron31.tokai.jaeri.go.jp}

\address[LNS]{
Istituto Nazionale Fisica Nucleare-Laboratorio Nazionale
del Sud, Via Santa Sofia 44, Catania 95123, Italy
}
\address[JAERI]{
Advanced Science Research Center, Japan Atomic Energy Research Institute,
Tokai, Ibaraki 319-1195, Japan
}
\address[CT]{
Istituto Nazionale Fisica Nucleare-Sezione di Catania,
Corso Italia 57, Catania 95129, Italy
}

\maketitle

\begin{abstract}
We investigate the formation and the decay of heavy systems
which are above the fission barrier.
By using a microscopic simulation of constraint molecular dynamics (CoMD)
on Au+Au collision,
we observe composite states stay for very long time 
before  decaying by fission.

\end{abstract}


\end{frontmatter}

The typical reaction mechanisms of heavy-ion collisions at lower 
incident energy are, depending on the energy and impact parameters, 
complete fusion, incomplete fusion, fusion-fission, molecular resonance,
and deep inelastic collisions.
Among the huge amount of studies in this field,
collisions of very heavy nuclei have been investigated
mainly for the creation of super heavy element (SHE).
SHEs are produced in two ways:
one is ``cold fusion'' which is complete fusion below the
classical barrier, and the other is ``hot fusion'' which
allows several neutrons to be emitted.
Even though the name is ``hot'', such reactions are still 
at very low energy near the barrier and the total mass number
is very close to the aimed one.
As far as the formation of SHE is concerned, the ``fusion'' of 
very heavy nuclei where the fission barrier no more exists is 
found to be ineffective \cite{Gaggeler,Herrmann}.

Apart from the formation of SHE, the study of fission dynamics,
including the spontaneous fission and the fusion-fission of heavy composite,
has been one of the most important subjects.
The competition of neutron emission between the fission
and the fission delay have been discussed intensively.
However almost all the discussion are done for mass regions 
where the classical fission barrier exists.

Sometime ago
many physicists paid attention to 
the low energy collision of very heavy nuclei 
in regard to the spontaneous positron emission from strong electric fields
\cite{SHM}.
If a molecule state of, say, U and U is formed and stays sufficiently 
long time, the binding energy of a electron can exceed the electron 
mass and might create electron-positron pair by a static QED process.
Unfortunately no clear evidence of static positron creation was observed 
below Coulomb energy region.
They have pointed out \cite{Reinhardt} the importance of nuclear 
reaction which causes the time delay of separation of two nuclei.
Although there increases
the background component of positrons from nuclear excitation,
which in this case is not interested in, 
the electron-positron from the static QED process 
is also expected to increase.
However, the reaction mechanism of very heavy nuclei has not been 
discussed by fully dynamical models.

In this paper we discuss the possibility of 
molecule-like states of heavy nuclei 
and the time scale of very heavy composite system formed
by the fusion-fission or deep inelastic processes.
To investigate these problems theoretically we use 
a recently developed constraint molecular dynamics (CoMD) model \cite{papa}.
This model has been proposed to include the Fermionic nature of
constituent nucleons by  a constraint that the phase space distribution 
should always satisfy the condition $f\leq 1$.
Among similar molecular dynamics models, 
there are quantum molecular dynamics (QMD)
\cite{refQMD}, Fermionic molecular dynamics (FMD) \cite{refFMD}, and
antisymmetrized molecular dynamics (AMD) \cite{refAMD}.
QMD has been the most popular and feasible model. 
Unfortunately it cannot, in principle, deal with the Fermionic 
nature of nuclear system, although sometime the 
phenomenological Pauli potential is introduced for such a purpose.
Therefore QMD model has been used mainly for higher energy phenomena
except for some exceptions \cite{maru90,refEQMD}.
More sophisticated models, i.e. FMD and AMD, deal with 
antisymmetrization of the wave function and have  
succeeded in describing nuclear reactions
at medium low energy and also in the study of nuclear structures.
However, due to the four dimensional matrix element of two-body interaction,
the CPU time  necessary to work out calculations for systems with
total mass larger than 200 is very large for practical studies.
The constraint molecular dynamics, on the other hand,
can deal to a certain extent with the Fermionic nature of the nuclear systems
and it is still feasible for heavy systems.

In this paper we apply CoMD to the investigation of 
$^{197}$Au+$^{197}$Au collisions at low energies where 
fusion-fission or deep-inelastic process may occur.
In the following we give a brief review of the model \cite{papa}.

The CoMD model mainly consists of two parts: classical equation of motion
of many-body system, and stochastic process which includes 
constraint of Pauli principle and the two-body collisions.
We write the distribution function of the system as a sum of
one-body distribution function neglecting the
antisymmetrization
\begin{eqnarray}
  f({\bf r},{\bf p}) &=&  \sum_i { f_i({\bf r},{\bf p}) },\\
f_i({\bf r},{\bf p}) &=&  \frac{1}{(2\pi\sigma_r\sigma_p)^3} \cdot
      \exp\left[-{({\bf r}-\langle{\bf r}_i\rangle)^2\over
2{\sigma_r}^2}
                -{({\bf p}-\langle{\bf p}_i\rangle)^2\over
2{\sigma_p}^2}\right].
\end{eqnarray}
The equation of motion of $\langle{\bf r}_i\rangle$ and $\langle{\bf
p}_i\rangle$
are derived using the time-dependent variational principle which gives:
\begin{equation}
\dot{\langle{\bf r}_i\rangle} =   \frac{\partial H}{\partial
\langle{\bf p}_i\rangle},
\;\;\;\;
\dot{\langle{\bf p}_i\rangle} = - \frac{\partial H}{\partial
\langle{\bf r}_i\rangle}.
\label{EOM}
\end{equation}
In our approach the total energy $H$ for $A$ particles with mass $m$
consists of
the kinetic energy and the effective interactions:
\begin{equation}
H=\sum_i {\langle{\bf p}_i\rangle^2\over 2m}+A{3\sigma_p^{2} \over 2m}+V
\label{hamiltonian}
\end{equation}
The second term arises from the Gaussian width in p-space.
However in the following considerations 
we omit such a constant term. 

The effective interaction $V$ we adopt is written as
\begin{equation}
V=V^{\rm vol}+V^{(3)}+V^{\rm sym}+V^{\rm surf}+V^{\rm Coul}.
 \label{potential}
\end{equation}
By defining the superimposition integral $\rho_{ij}$ as:
\begin{eqnarray}
\rho_{ij} &\equiv&
   \int d^3r_i\;d^3r_j\; \rho_i({\bf r}_i)\rho_j({\bf r}_j)
\delta({\bf r}_i-{\bf r}_j), \\
\rho_i &\equiv& \int d^3p\;f_i({\bf r,p}),
\end{eqnarray}
the terms in Eq.~(\ref{potential}) can be written as:
\begin{eqnarray}
V^{\rm vol}&=&{t_{0} \over 2\rho_{0}}\sum_{i,j\neq i}\rho_{ij}, \\
V^{(3)}&=&{t_{3} \over (\mu+1)(\rho_{0})^{\mu}}\sum_{i,j\neq
i}\rho_{ij}^{\mu}, \\
V^{\rm sym}&=&{a_{\rm sym} \over 2\rho_{0}}\sum_{i,j\neq i}
 [2\delta_{\tau_i, \tau_j}-1]\rho_{ij}, \\
V^{\rm surf}&=&{C_{s} \over 2\rho_{0}}\sum_{i,j\neq i}
\nabla^{2}_{\langle{\bf r}_{i}\rangle}(\rho_{ij}), \\
V^{\rm Coul}&=&{1\over2}\sum_{i,j\neq i\atop (i,j\in {\rm protons})}
{e^2 \over |\langle{\bf r}_{i}\rangle-\langle{\bf r}_{j}\rangle|}
{\rm erf}\left({|\langle{\bf r}_{i}\rangle-\langle{\bf r}_{j}\rangle|
 \over 2\sigma_{r}^{2}}\right).
\end{eqnarray}
In the above relations 
the coordinate $\tau_{i}$ represents the nucleon 
isospin degree of freedom.

We have two sets of parameters, 
regarding the interaction strength
and the width of distribution function,
which are different mainly for the stability of the system.

The parameter set (I) used in \cite{papa}, 
$\sigma_{r}=1.3$ fm,
$\sigma_{p}/\hbar=0.47$ fm$^{-1}$, 
$t_{0}=-356$ MeV,
$t_{3}=303$ MeV,
$\mu=7/6$,
$a_{\rm sym}=32$ MeV,
$C_{s}=-0.33$ MeV$\cdot$fm$^{2}$,
$\rho_0=0.165$ fm$^{-3}$, gives a good reproduction of fragmentation 
data on Ca+Ca and Au+Au at 35 MeV/nucleon, 
and the mean radii and the binding energies in a wide range of mass.

The parameter set (II), which we introduce in this paper,
($\sigma_{r}=1.15$ fm,
$\sigma_{p}/\hbar=0.4748$ fm$^{-1}$, 
$t_{0}=-301.1$ MeV,
$t_{3}=242$ MeV,
$\mu=7/6$,
$a_{\rm sym}=26.4$ MeV,
$C_{s}=-0.165$ MeV$\cdot$fm$^{2}$,
and $\rho_0=0.165$ fm$^{-3}$), reproduces reasonably well the fusion 
cross section of Ca+Ca reactions,
while set (I) overestimates such a data.
Even though we have not been able yet to find a unique parameter set 
consistent with both features of fusion and fragmentation, 
we find some experimental confirmations of our calculations
as mentioned above.  
In this work, since we apply the model to an energy region and 
to very heavy systems for which the experimental data is scarce, 
we plan to give upper and lower estimates which will be interesting 
to confirm experimentally.
We further strengthen our results with Boltzmann-Nordheim-Vlasov (BNV) 
calculations\cite{BNV}. 

The Pauli principle is taken into account in two ways:
One is the Pauli blocking of the final state of two-body collision
and the other is the constraint which brings into the system the 
Fermi motion in a stochastic way.
The starting point of the constraint is the requirement:
\begin{eqnarray}
\overline{f}_{i} &\leq& 1\ \ \ \ \ \ \hbox{(for all $i$)}, \label{constraint}
\\
\overline{f}_{i} &\equiv&
 \sum_j \delta_{\tau_{i},\tau_{j}}
  \delta_{s_{i},s_{j}}
  \int_{h^{3}} f_j({\bf r}, {\bf p})\;d^3r\;d^3p, \label{occupation}
\end{eqnarray} 
where $s_i$ is the spin coordinate of the nucleon $i$.
The integral is performed in an hypercube
of volume $h^{3}$ in the phase-space
centered around the point $(\langle {\bf r}_i\rangle,\langle {\bf
p}_i\rangle)$
with size
$\sqrt{{2\pi\hbar \over \sigma_{r}\sigma_{p}}}\sigma_{r}$ and
$\sqrt{{2\pi\hbar \over \sigma_{r}\sigma_{p}}}\sigma_{p}$ in the
$r$ and $p$ spaces respectively.

At each time step and for each particle $i$ 
the phase space occupation $\overline{f}_i$ is checked.
If $\overline{f}_i$  has a value greater than 1
an ensemble $K_i$ of nearest particles (including the particle $i$) 
is determined within the distances $3\sigma_r$ and $3\sigma_p$ 
in the phase space.
Then we change randomly the momenta of the particles belonging to
the ensemble $K_i$ in such a way that for the newly generated sample 
the total momentum and the total kinetic energy is conserved
(``many-body elastic scattering'').
The new sample is accepted only if it
reduces the phase space occupation $\overline{f}_i$\cite{papa}.

To handle the Pauli-blocking in the collision term is straightforward
from the constraint.
In fact for each NN collision we evaluate the occupation probability
after the elastic scattering.
If such functions are both less than 1 the collision
is accepted, rejected otherwise.   
We note that for the results discussed here and especially at 
the lowest energies the collision term is of little importance.

To simulate the collision of two $^{197}$Au nuclei, 
we prepare the ground state by applying the frictional cooling
method together with the constraint of CoMD.
The ground states we obtain have
binding energy of 7.6 MeV/nucleon and a root mean square radius of 5.76 fm 
with parameter set (I)
and 8.4 MeV/nucleon and 5.34 fm for parameter set (II).
They are rather stable for 1000 fm/$c$. 
For instance our $^{197}$Au ground states 
with parameter sets (I) and (II) evaporates 2.75 and 3.1 
nucleons during 1000 fm/$c$, respectively.
The collision events are performed for impact parameter $b$ of 0 and 6 fm 
for incident energy in laboratory system of $E_{\rm lab}=5\sim35$ MeV/nucleon.

 Figure~1 shows a typical event of CoMD (I) calculation 
with incident energy $E_{\rm lab}=10$ MeV/nucleon  
with impact parameter $b=6$ fm. The
two nuclei form a quite deformed compound system, they keep such a  
deformation almost 2500 fm/$c$ and finally fission takes place.
The system does not show much rotation since the angular momentum per nucleon
is not so large and the elongated shape makes the moment of inertia larger 
than that in the initial stage.
Therefore the reaction mechanism we are observing here may be 
in-between the deep inelastic and molecular resonance.

\begin{figure}
\epsfxsize=0.9\textwidth
\epsfbox{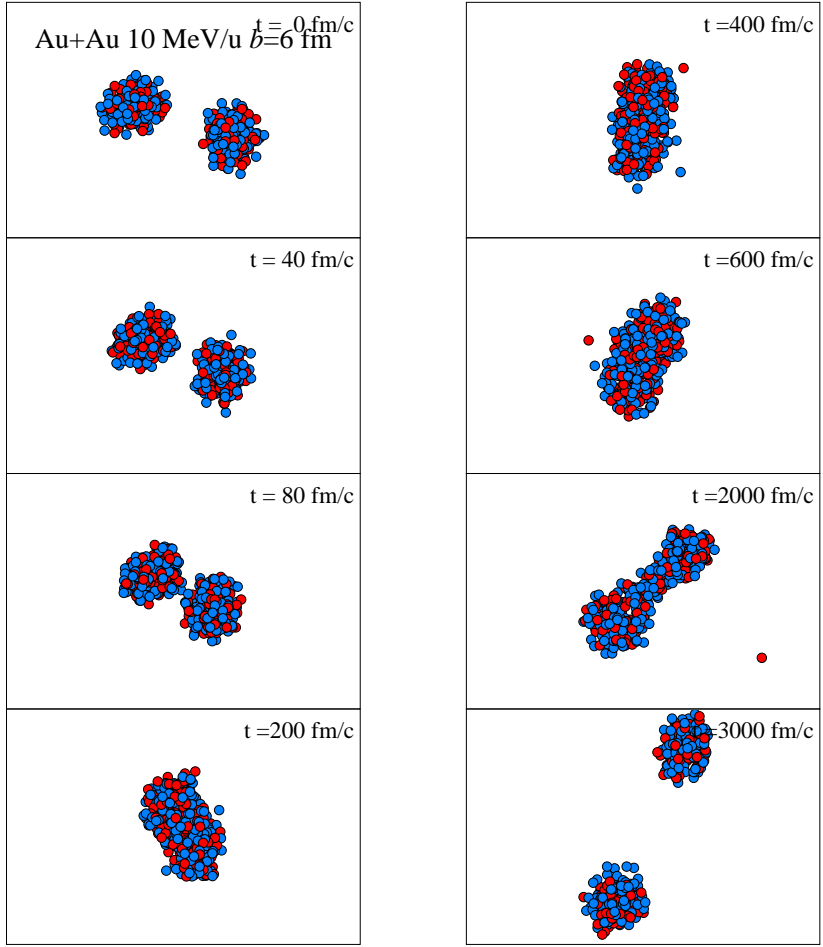}
\caption{
Snapshot of $^{197}$Au+$^{197}$Au at $E_{\rm lab}=10$ MeV/nucleon $b=6$ fm.
The time indicated in each panel is not from the contact of 
two nuclei but indicates only that of the simulation.
}
\end{figure}

There are many observables which distinguish the reaction mechanism.
The largest fragment mass is one of such well-defined 
observables which can easily be measured experimentally.
Figure 2 shows the time dependences of the largest cluster mass 
for the impact parameters $b=0$ and 6 fm 
calculated by CoMD (I), CoMD (II),  QMD  and BNV.
In CoMD calculations we see at the beginning the largest cluster mass 
$A_{\rm max}=197$ which corresponds to projectile and target mass number.
Within about 50 fm/$c$, $A_{\rm max}$ becomes 394 
except for the incident energy $E_{\rm lab}=5$ MeV which is 
below the barrier where two nuclei never contact.
At incident energies above the barrier, the formed large system
will decay into smaller fragments by different modes according to
the energy and angular momentum.
At higher incident energies ($E_{\rm lab}\geq 30$ MeV/nucleon)
the largest cluster mass changes suddenly at the early stage
and continuously decreases in time. 
This indicates multifragmentation for head-on collisions 
and deep inelastic reaction for peripheral collisions followed by 
the emission of nucleons and small fragments.
At lower incident energies ($E_{\rm lab}\leq 20$ MeV/nucleon) 
there is a sudden change of the largest cluster mass 
at very late time, which indicates a fission of the system. 
One should note that in our calculation of Au+Au system
there is almost no event where the system decays only by 
emitting particles or light fragments, i.e., pure incomplete fusion.
The instability due to the Coulomb repulsion plays the major role
in the decay process. 

\begin{figure}
\epsfxsize=0.9\textwidth
\epsfbox{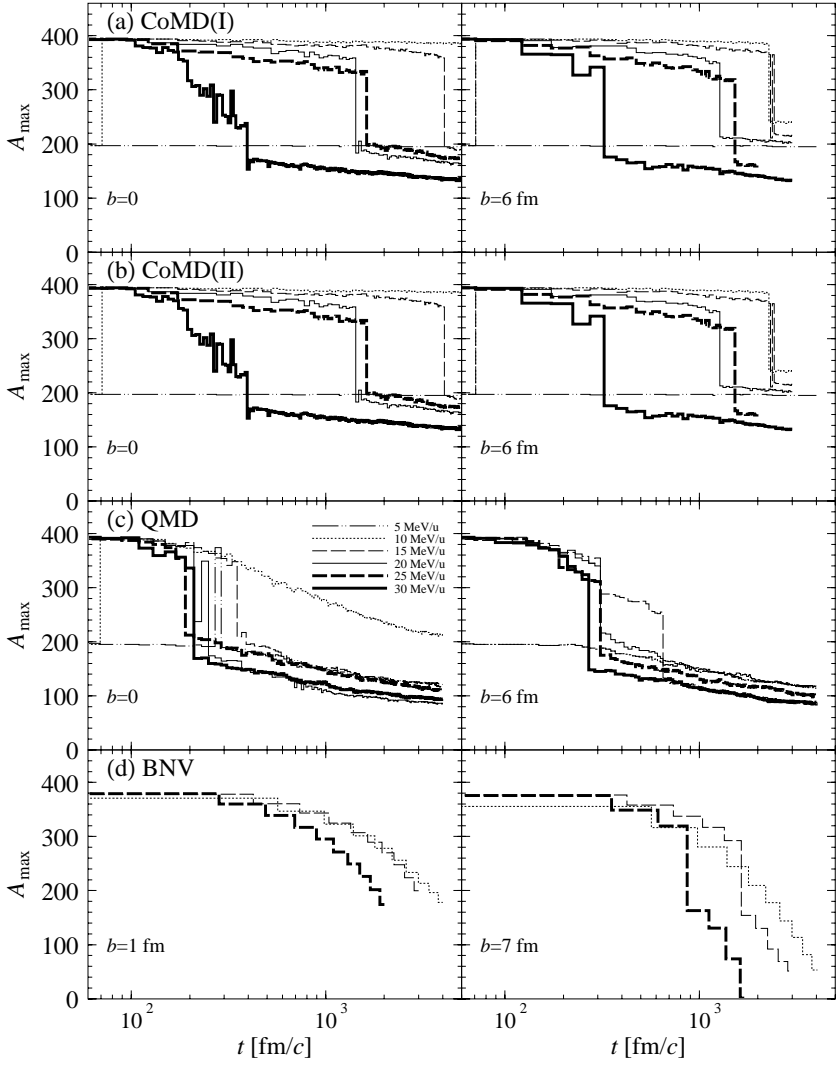}
\caption{
The time-dependence of the largest fragment mass $A_{\rm max}$.
 From the top (a) CoMD with parameter set (I), (b) CoMD (II), 
and (c) QMD.
The left panels show cases of head-on collision and the right $b=6$ fm. 
The lowest panels (d) refer to BNV calculations at 1 and 7 fm respectively.
}
\end{figure}

Here we should note that the plotted largest fragment mass are obtained by 
only one event for each incident energy and impact parameter.
Therefore the fission time includes large amount of statistical error.
In fact the case of $E_{\rm lab}=10$ MeV/nucleon with $b=0$ fm,
fission process is not observed in Fig.~2 (a).
With a different series of initial conditions, however,
we observe fission of the system around $t=10^{4}$ fm/$c$ for 
parameter (I).
In Fig.~2 (c) the same quantity as Fig.~2 (a) and (b) is displayed 
for QMD calculation.
These QMD calculations are based on the same code as CoMD (I) 
switching off the constraint procedure.
The difference between the CoMD and the QMD is clear and dramatic.
At low energy collisions there are no fission process 
and the system decays only by emitting nucleons and light fragments.
At higher energies there is some sudden change of the largest fragment mass
even in QMD calculation. 
This is not a fission but passing through for head-on collisions
or deep inelastic process for peripheral collisions.

In BNV \cite{BNV} calculation (Fig.~2 (d)) the sudden change of
maximum fragment mass number is not observed for impact parameter $b=1$ fm
and $b=7$ fm at low energy,
except for $b=7$ fm and $E_{\rm lab}\geq 15$ MeV/nucleon.
For the fission process with very small angular momentum, 
 fluctuations and correlations are very important which are not
included in the BNV calculation.
Instead, the system decays via evaporation of nucleons like the QMD case.
One should note, however, that the Pauli principle is satisfied in BNV 
calculation while it is not in QMD case.
The time scale of the very large composite is still of the same order
as in CoMD (II) calculation.
Although the reaction mechanism is different for CoMD and BNV,
this similarity of time scale supports the validity of our CoMD calculation.

Assuming a very simple form of the time-dependent fission 
width $\Gamma(t)=\Gamma_{\rm f}\;\theta(t-T_{\rm d})$,
the averaged fission time $T_{\rm fiss}$ can be obtained by the 
survival probability of the compound system
against two-body process $P_{\rm surv}$ as
\begin{eqnarray}
P_{\rm surv} &=& \exp\left[-(t-T_{\rm d})\Gamma_{\rm f}/\hbar \right], 
\label{Tfission1}\\
T_{\rm fiss} &\equiv& T_{\rm d}+\hbar/\Gamma_{\rm f} 
\label{Tfission2}
\end{eqnarray}
where $T_{\rm d}$ is the delay time and $\Gamma_{\rm f}$ is the 
``fission width'' after the delay time.
The probability $P_{\rm surv}(t)$ is obtained directly by the simulation.
This fitting can apply well only for fission-like process 
in our calculation.
Figure 3 shows the survival probability of a large fragment with $A>350$ 
which decays by fission-like mode or emitting large fragments ($A>30$).
The histograms are directly obtained by the simulation and the curves
are fitting  by Eqs.~(\ref{Tfission1}-\ref{Tfission2}).
 From the top, results of CoMD with parameter set (I), CoMD (II) and QMD
are listed for impact parameter $b=0$ and for several incident energies
$E_{\rm lab}=10 \sim 25$ MeV/nucleon.
For all the calculations the fitting works well, particularly 
the effect of delay time.
The assumption of constant fission width after the delay time,
on the other hand, is not completely supported 
because of poor statistics and still existing dynamical effects.
One should note that the fitting by Eqs.~(\ref{Tfission1}-\ref{Tfission2})
is just to extract the ``fission'' time of the super heavy composite.
Especially the time scale of QMD results is obviously different from
that of fission process.

\begin{figure}
\epsfxsize=0.5\textwidth
\epsfbox{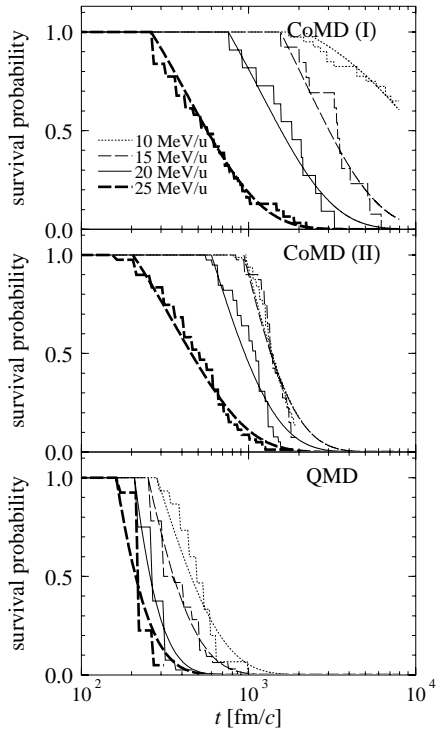}
\caption{
The survival probability of large fragments ($A>350$)
which decay by fission-like mode.
The abscissa indicates the time after the contact of two nuclei.
 From the top the results are obtained for head-on collisions 
by CoMD with parameter set (I), CoMD (II), and QMD.
The histograms indicate results from simulations 
and smooth curves are the fits by Eqs.~(15-16).
}
\end{figure}

The extracted fission time $T_{\rm fiss}$ are plotted in Fig.~4.
The fission time shown for parameter set (I) 
might be too long for such heavy system.
To make more quantitative discussions we should improve 
the effective interaction.  
By using parameter set (II), we obtain smaller values of fission time. 
We can consider the values obtained as upper- and lower-limits of 
the fission time in our CoMD model.
However the experimental data will finally support one or 
the other result which, we stress, are both qualitatively 
similar and somehow surprising.
For both of parameter sets the longest life time of very heavy composite 
is found at $E_{\rm lab}=10$ MeV/nucleon.

For lower incident energies (just above the Coulomb barrier) 
the system might not form a fully thermalized single composite but might be 
quasi separated in the phase space, which makes the system split easily. 
For higher energies, the fully thermalized system needs 
some fluctuations to reseparate even though there is no classical
barrier for fission.
Therefore the fission time gets shorter with increase of the incident energy.

\begin{figure}
\epsfxsize=0.5\textwidth
\epsfbox{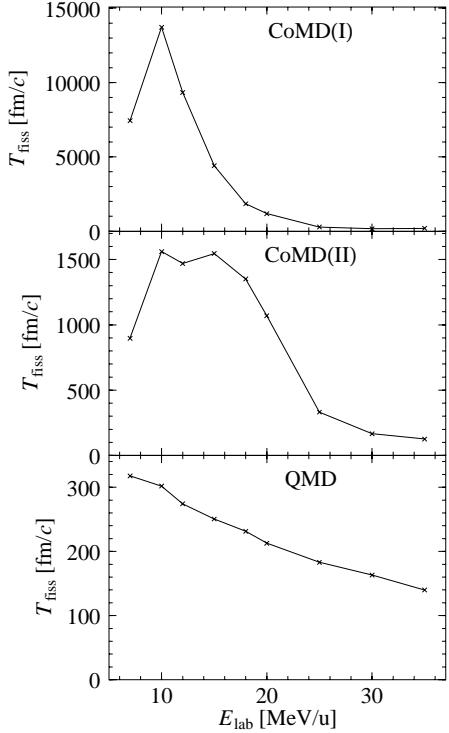}
\caption{
The fission life time obtained from the Eqs.~(15-16).
}
\end{figure}

In QMD calculation what is in marked contrast to the CoMD calculation is
that the ``fission time'' has no maximum energy and shows monotonic decrease. 
This is due to the lack of the Pauli principle which suppresses
two nuclei from overlapping at very low energies (above the Coulomb barrier).

For peripheral collision ($b=6$ fm), the life time of very heavy 
composite is  shorter than the head-on collisions.
But the incident-energy dependence is very similar to the $b=0$ fm cases.
Though the mechanism is much more dynamical,
Eqs.~(\ref{Tfission1}-\ref{Tfission2}) fit well again.

{\it Nevertheless, the super heavy composite system 
formed by the head-on collision 
of Au+Au may survive rather long time of $10^3 \sim 10^4$ fm/$c$.}
We note that such long-lived strongly deformed (see Fig.~1) systems 
have been observed by looking at the binary dissipative collisions
between lighter system ($A_{\rm total}\simeq 60$) in the same scaled 
energy regime with respect to the Coulomb barrier. 
This long time intervals have been well estimated through the comparison
of the incident-energy averaged
angular distributions and/or the excitation functions with the results
of the partially overlapped molecular level model (POMLM)\cite{papa1}.
Such studies could be surely extended, at least concerning
the average angular distribution of the binary processes,
also in the present case.

Moreover, another interesting aspect of
the long-lived very heavy system is, as mentioned before, 
the spontaneous positron-electron production from the strong 
electric field as a static QED process.
The total charge of Au+Au system may be still smaller than the
necessary charge ($Z\sim 170$) for this process.
However, the nuclear reaction of, e.g. U+U system, should be 
qualitatively the same as what we observe in Au+Au system.
Although the background positrons should be larger,
one can get longest life time of strong electric field 
(stronger than the case of Rutherford or molecular trajectory)
around $E_{\rm lab}=10$ MeV/nucleon at some impact parameter
and the production of positrons from the static QED process
should be largest around that energy.

As mentioned above, production of SHE is one of the most important
subject in the heavy-ion collision problem.
Besides cold- and hot-fusion, mass transfer in collision of 
very heavy nuclei was tried before.
One could produce, e.g. up to Fm ($Z=100$) in U+U system,
or Md ($Z=101$) in U+Cm system,
by such a mechanism \cite{Gaggeler,Herrmann}.
The incident energy, however, was very close to the Coulomb barrier
and the reaction was rather gentle with the transfer of $\sim 20$ nucleons.
In our CoMD calculation for $E_{\rm lab}\geq 7$ MeV/nucleon,
the reaction mechanism is more violent and
there happens the transfer of much more nucleons though
the mass loss from the system is also large.
In Fig.~5 plotted is the mass-asymmetry 
$(A_1-A_2)/(A_1+A_2)$
of the fission process in CoMD calculation,
where $A_1$ and $A_2$ are the largest and the second largest
fragment mass when the fission occurs.
The mass-asymmetry increases with the incident energy.
At $E_{\rm lab}=7$ MeV/nucleon, the asymmetry amounts to
about 0.1 
and at 10 MeV/nucleon almost 0.2 as average.
If we simply assume no proton loss and asymmetry of 0.2
the largest fragment charge will be 112 for U+U system.
Of course we should consider the thermal mass loss and
subsequent fission due to the excitation of fragments.
However, such kind of fusion-fission mechanism at around
10 MeV/nucleon should be taken into account 
for the SHE production.
The new 4$\pi$ detectors can accumulate lots of statistics plus 
they can make coincidence studies to see if the fragments come from fission.

\begin{figure}
\epsfxsize=0.5\textwidth
\epsfbox{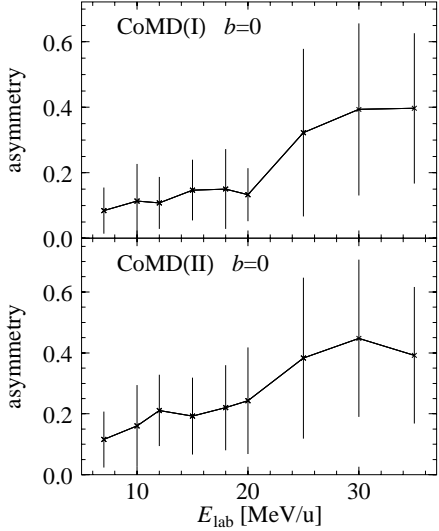}
\caption{
The mass asymmetry of the fission fragments.
Error bars indicate statistical standard deviation.
}
\end{figure}

In summary, we have discussed the formation and  decay of
super heavy composite in the Au+Au collisions.
The CoMD calculation
which takes into account the Fermionic nature of the nucleon many-body
system can describe well the low-energy dynamics including
fusion, fission, deep inelastic, emission of nucleons and small fragments,
and multifragmentation.
Although there are still some ambiguities on the effective interaction,
the life time of super heavy composite is found to be rather long
up to $10^3\sim10^4$ fm/$c$.
Some experimental explorations such as detection of $e^+e^-$ formation 
at around 10 MeV/nucleon 
and measurement of the energy averaged angular distribution and/or
excitation function for binary processes
are encouraged.

One of the authors T.M. thanks INFN-LNS for warm hospitality during
his stay and Dr.~A.~Iwamoto, Dr.~H.~Ikezoe, and Dr.~S.~Mitsuoka 
for fruitful discussions. A.B. thanks Prof.~J.~Natowitz for enlighting 
discussion on the super heavy system discussed here.

\end{document}